\pdfoutput=1
\documentclass[11pt]{article}

\usepackage{acl}

\usepackage{times}
\usepackage{latexsym}

\usepackage[T1]{fontenc}
\usepackage[utf8]{inputenc}

\usepackage{microtype}
\usepackage{todonotes} 
\usepackage{braket}
\usepackage{amsmath}
\usepackage{amsfonts}

\title{Quantum Natural Language Generation on Near-Term Devices}

\author{Amin Karamlou\thanks{$\;$ corresponding author: Amin.Karamlou@cs.ox.ac.uk.} \\ IBM Quantum \\ University of Oxford \\
        \And  Marcel Pfaffhauser \\ IBM Quantum \And
        James Wootton \\ IBM Quantum}
\begin{document}
\maketitle
\begin{abstract}
The emergence of noisy medium-scale quantum devices has led to proof-of-concept applications for quantum computing in various domains. Examples include Natural Language Processing (NLP) where sentence classification experiments have been carried out, as well as procedural generation, where tasks such as geopolitical map creation, and image manipulation have been performed. We explore applications at the intersection of these two areas by designing a hybrid quantum-classical algorithm for sentence generation. 

Our algorithm is based on the well-known simulated annealing technique for combinatorial optimisation. An implementation is provided and used to demonstrate successful sentence generation on both simulated and real quantum hardware. A variant of our algorithm can also be used for music generation.

This paper aims to be self-contained, introducing all the necessary background on NLP and quantum computing along the way.
\end{abstract}

\section{Introduction}
It is widely believed that computers operating according to the laws of quantum mechanics will outperform classical computers at specialised tasks. This belief is backed up by the fact that important computational problems such as integer factorisation \cite{Shor1997factoring} and unstructured search \cite{Grover1996search} admit quantum algorithms which are provably faster than the best known classical algorithms for solving them. Unfortunately, in order to make use of these algorithms, we would first need to build scalable, fault-tolerant quantum computers, which are still some years away. By contrast, the current generation of quantum computers are still fairly rudimentary, containing at most a few hundred noisy qubits, i.e. qubits with which we cannot perform perfect operations~\cite{Preskill2018nisq}. Despite their shortcomings, these devices represent a significant milestone for quantum computing. This is because unlike their smaller predecessors they cannot be simulated efficiently on classical hardware. Hence, it is possible that near-term quantum devices will bring with them the first examples of tasks performed by quantum computers that not even the most powerful classical supercomputers can perform, with tentative first steps made for proof-of-principle problems \cite{arute2019supremacy,pednault2019summit}. The search for examples in which a useful advantage can be demonstrated has led to the development of tailor-made algorithms for near-term devices that solve problems in domains such as chemistry, and optimisation \cite{farhi2014quantum, peruzzo2014variational}.

In this paper, we are concerned with near-term quantum algorithms for natural language generation (NLG). NLG lies at the intersection of procedural generation, i.e. the algorithmic generation of data, and Natural Language Processing (NLP), both of which are active research topics within the quantum software community (see e.g. \citealp{wootton2020map_gen, wootton2020proc_gen, coecke2020foundations_qnlp, lorenz2021qnlp}). The importance of NLG is underscored by its wide range of potential applications. It can for instance be used in video games to create natural-sounding dialogue, or in journalism to create automated news articles. These applications are often time-sensitive, as in the case of video games, where delays in dialogue generation would make the user experience unsatisfactory. In other situations, NLG algorithms have to deal with a large amount of input data. This is the case in automated journalism where information from many different sources needs to be collated into one coherent article. These considerations mean that developing faster algorithms for NLG tasks would have tremendous practical consequences. Thus, it is natural to wonder if any such tasks can benefit from speedups when performed on a quantum computer. Our aim here is to take the first steps towards answering this question.

Throughout this work, we will make use of the well-established mathematical connection between the Distributional Compositional Categorical (DisCoCat) \cite{Coecke2010DisCoCat} model of natural language and quantum theory. This connection was recently exploited in several works \cite{meichanetzidis2020grammar_aware, lorenz2021qnlp} to successfully perform Quantum Natural Language Processing (QNLP) on real quantum hardware (as opposed to simulation with conventional hardware). More specifically it was used to perform the task of binary sentence classification. The aim of this task is simple: Given a sentence about one of two possible topics, decide which topic it is about. Building upon this work, we design a sentence generation algorithm that can run on current quantum hardware. Our algorithm takes as input one of several possible topics and produces as output a sentence with that topic. Our algorithm works by searching through the space of possible sentences using simulated annealing (SA), a well-known probabilistic method for solving combinatorial optimisation problems. The choice of SA is motivated by the recent success of the method at (classically) solving the task of sentence paraphrasing \cite{liu2020annealing}. We experimentally evaluate the performance of our algorithm at news headline generation. We also show how our algorithm can be adapted to perform music generation.

Before continuing it is worth clarifying the goal of this paper and the scope of our claims. The formal similarity between DisCoCat and quantum theory has led to some authors claiming that NLP is an inherently ``quantum native`` field \cite{coecke2020foundations_qnlp}, and that we can expect large-scale quantum computational speedups for NLP tasks as more powerful quantum hardware becomes available. Testing these claims theoretically would require significant analysis of QNLP proposals using computational complexity theory, as has been done with other proposals for quantum advantage, for example in \citet{aaronson2016complexity, brakerski2020simpler, zhu2021interactive}. Alternatively, we could wait for larger quantum computers to be built, allowing for experimental comparison of QNLP algorithms and cutting-edge classical methods such as GPT-3 \cite{Brown2020gpt-3} or BERT \cite{Devlin2019bert}. We do not claim to address either one of these challenges here. Our work is rather a proof-of-concept example of how NLG can be performed on quantum hardware. We also hope that by assuming a modest mathematical background this paper can serve as an introduction to quantum software design using the diagrammatic style of quantum theory utilised in QNLP research.

The rest of the paper is organised as follows: In section \ref{section:prelims} we describe the necessary background on DisCoCat and quantum computing. Section \ref{section:algorithm} contains the details of our SA-based sentence generation algorithm. We report the results of experiments with this algorithm in section \ref{section:experiments}, including a discussion of how the algorithm can be adapted for music composition in section \ref{section:music}. Finally, we discuss future research avenues in section \ref{section:future_work}.

\section{Preliminaries}
\label{section:prelims}
\subsection{Quantum Computing}
This section presents a self-contained overview of the basics of quantum computation, assuming no familiarity with the topic. Naturally, what we present is far from a complete introduction. A more in-depth book for further reading is \citet{nielsen2002quantum}. Alternatively, \citet{coecke2018picturing} introduces quantum theory via the diagrammatic language used here.

The idea behind quantum computation is to harness features of quantum mechanics that have no classical analogue in the design of efficient algorithms. The first of these features worth mentioning is called \emph{superposition}. The logical building blocks of a classical computer are bits. These are objects that can have one of two possible states, 0 or 1. The quantum analogue of a bit, known as a qubit, has a state that lives in a 2-dimensional Hilbert space. We use the notation\footnote{This is referred to as Dirac or bra-ket notation and is used ubiquitously throughout quantum information. See appendix 10 of \cite{deWolf2019notes} for a concise introduction to this formalism.} $\ket{0} = \begin{bmatrix}1\\0\end{bmatrix}$ and $\ket{1} = \begin{bmatrix}0\\1\end{bmatrix}$ to denote the orthonormal basis vectors of this space. The state of a qubit, written as $\ket{\psi}$, is a linear combination of these basis vectors:
$$\ket{\psi} = \alpha \ket{0} + \beta \ket{1} \; \text{s.t.} \; \alpha, \beta \in \mathbb{C}, |\alpha|^2 + |\beta|^2 = 1$$
It is this linear combination that is referred to as a superposition.

The act of reading the value of a qubit in state $\ket{\psi}$ is called a \emph{measurement}. Regardless of what superposition a qubit is in, the result of a measurement is always one of two possible outcomes, $0$ or $1$. The probability of measuring $0$ is equal to $|\alpha|^2$, and $\alpha$ is known as the \emph{amplitude} of $\ket{0}$. Likewise, the probability of measuring $1$ is $|\beta|^2$, and $\beta$ is known as the \emph{amplitude} of $\ket{1}$. Crucially, once a measurement has occured, the state $\ket{\psi}$ \emph{collapses} to the corresponding basis state. For example, if we measure a qubit in state $\ket{\psi} = \frac{1}{\sqrt{2}} \ket{0} + \frac{1}{\sqrt{2}} \ket{1}$ and observe the outcome 0, then immediately after the measurement the state of the qubit is $\ket{0}$.

Naturally, to perform a meaningful computation we need to use more than just one qubit. The \textit{joint state} $\ket{\phi}$ of $n$ qubits lives in a Hilbert space of dimension $N = 2^n$ with orthogonal basis states of the form $\ket{b_1} \otimes \ket{b_2} \otimes ... \otimes \ket{b_n}$ where each $b_i \in \{0, 1\}$. We will abbreviate these basis states to $\ket{b_1 b_2b_3 ...b_n}$. With some abuse of notation it will also often be convenient to write these basis states in decimal notation i.e. $\ket{0} = \ket{000...000}, \ket{1} = \ket{000...001}, \ket{2} = \ket{000...010}, ... \ket{N-1} = \ket{111...111}$.

$\ket{\phi}$ is then once again a superposition: 

\begin{align*}
&\ket{\phi} = \alpha_0 \ket{0} + \alpha_1 \ket{1} + ... \alpha_{N-1} \ket{N-1} \\
& \text{s.t.} \; \forall{i} \alpha_i, \beta \in \mathbb{C}, \sum_i|\alpha_i|^2 = 1
\end{align*}

When measuring $\ket{\phi}$ one observes outcome $i$ with probability $|\alpha_i|^2$ and the state of the underlying qubits collapses to $\ket{i}$.

Aside from measurement, a quantum system can also be manipulated using \emph{quantum logic gates}. Mathematically, these gates are unitary linear maps $U$. Thus, the evolution of a system from one timestamp to the next can simply be described as $\ket{\psi_1} = U \ket{\psi_0}$.

\begin{figure}
    \centering
    \includegraphics[width=0.9\textwidth /2]{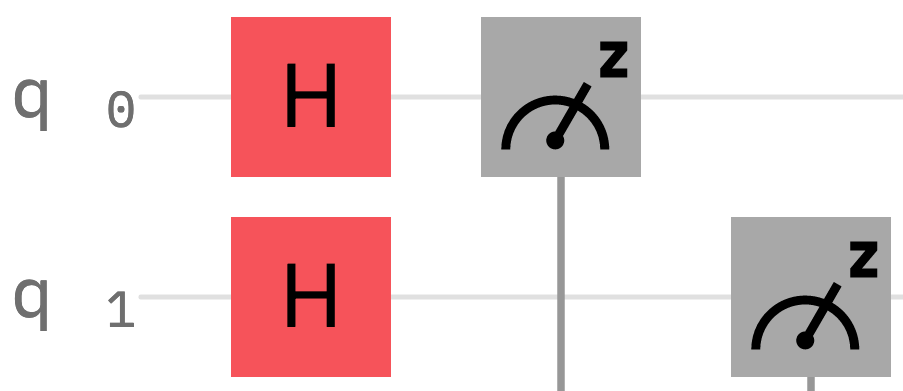}

    \caption{A simple quantum circuit created using the IBM Quantum Composer available at \url{https://quantum-computing.ibm.com/.}}
    \label{fig:quantum_circuit}
\end{figure}

Pictorially, a quantum computation can be represented as a circuit. Figure \ref{fig:quantum_circuit} provides an example of such a circuit. In this example, two qubits begin in the joint state $\ket{\psi_0} = \ket{0}$. A quantum logic gate $H = \frac{1}{\sqrt{2}} \begin{bmatrix}
1 & 1\\
1 & -1
\end{bmatrix}$, known as a hadamard gate is applied to each qubit, transforming the state into $\ket{\psi_1} = H \otimes H \ket{0} = \frac{1}{2} \ket{0} + \frac{1}{2} \ket{1} + \frac{1}{2} \ket{2} + \frac{1}{2} \ket{3}$. Finally, the state is measured, resulting in one of the four possible outputs $0, 1, 2$, or $3$ being observed, each with a probability of $\frac{1}{4}$. After measurement, the state collapses to the respective basis state $\ket{0}, \ket{1}, \ket{2}$, or $\ket{3}$.

\subsection{DisCoCat and QNLP}

The Distributional Compositional Categorical (DisCoCat) model of language meaning \cite{Coecke2010DisCoCat} is a mathematical framework that allows for the meaning of a sentence to be described as a combination of the meaning of its constituent words, and the grammatical relationships between these words. This is in contrast to many older NLP models, which treat sentences as ``bags of words'' while ignoring their grammatical structure.

\begin{figure}
    \centering
    \includegraphics[width=\textwidth /2]{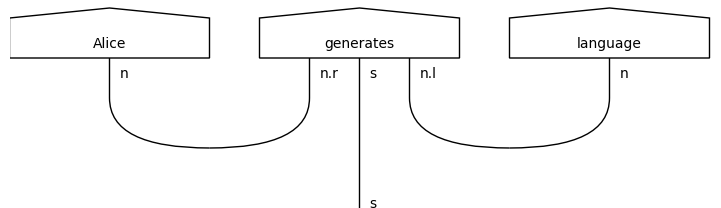}
    \caption{DisCoCat diagram for the sentence `Alice generates language.'}
\label{fig:discocat_diagram}
\end{figure}

DisCoCat comes equipped with a pictorial representation, allowing any sentence to be represented by a so-called \emph{string diagram}. Such a diagram consists of boxes representing words, and wires connecting these boxes according to the formalism of pregroup grammars \cite{lambek2008word}. This means that every wire in the diagram is annotated either by some atomic type $p$, a left adjoint $p.l$, or a right adjoint $p.r$. Let us explain the role of types and adjoints through example, by considering the sentence ``Alice generates language``. The DisCoCat diagram corresponding to this sentence is given in figure \ref{fig:discocat_diagram}. In this diagram, wires are annotated by the noun type $n$ and the sentence type $s$. As we can see, the box for the word `generates' has three wires coming out of it, which are annotated by $n.r, s$, and $n.l$ respectively. This indicates that the word `generates' expects to receive a noun on its left (in this case `Alice'), as well as another noun on its right (in this case `language') in order to output a grammatical sentence. In general, a sentence is grammatical if its DisCoCat diagram has a single open output wire of type $s$, as in the example of figure \ref{fig:discocat_diagram}.

It is worth noting that DisCoCat diagrams are more than simple pictures. They are based on the rigorous formalism of monoidal categories \cite[Chapter~1]{heunen2019categories}, which means they are equipped with a diagrammatic calculus. This calculus can be used to rewrite complicated string diagrams into simpler ones that still encode the meaning of the original sentence. As it happens, monoidal categories and string diagrams also turn out to be a suitable high-level framework for capturing much of quantum information and computation \cite{abramsky2004categorical, coecke2018picturing}. This observation is part of the reason that one may hope for quantum advantage in NLP tasks in the long term. 

We now outline a procedure for transforming any sentence into a parameterised quantum circuit that can be run on real IBM Quantum hardware. The pipeline we discuss here has recently been implemented as part of \emph{lambeq} \cite{kartsaklis2021lambeq}, a python library developed specifically for QNLP tasks.

\begin{enumerate}
    \item A sentence is converted to a DisCoCat diagram using the Combinatory Categorical Grammar (CCG) based techniques of \citet{yeung2021ccgbased}.
    \item The DisCoCat diagram is simplified using some of the rewrite rules available in lambeq. Even though this step is strictly speaking optional, applying rewrite rules often leads to crucial computational advantages, for instance by reducing the number of qubits required to implement the parameterised quantum circuit.
    \item An \emph{ansatz} is used to transform the simplified diagram to a parameterised quantum circuit. This ansatz is a mapping that assigns a number of qubits to each wire type in the string diagram, as well as a set of quantum logic gates to each word in the diagram.
    \item The quantum compiler t$\ket{\text{ket}}$ \cite{Sivarajah2020tket} is used to translate the parameterised quantum circuit into machine-specific instructions, which can be executed on real IBM quantum computers.
\end{enumerate}

\begin{figure*}
    \centering
    \includegraphics[width=\textwidth]{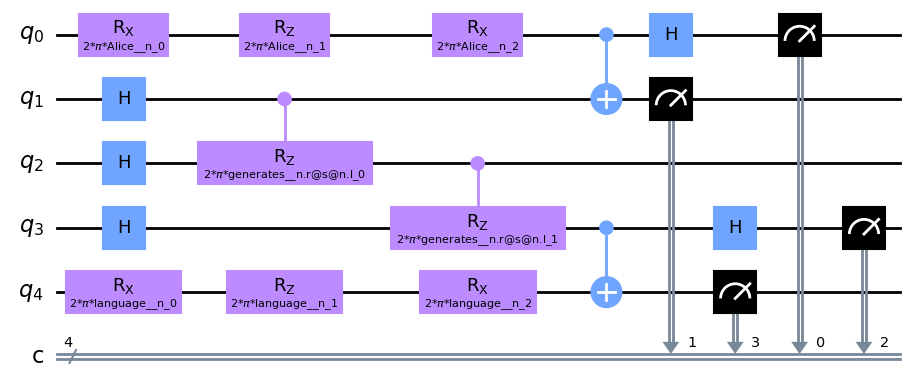}
    \caption{Parameterised quantum circuit for the sentence ``Alice generates language''.}
\label{fig:qiskit}
\end{figure*}

In this paper we use the IQP ansatz. This transforms each DisCoCat diagram into an Insantanoues Quantum Polynomial (IQP) circuit. We do not justify this choice of ansatz here, more information is available in \cite{havlicek_supervised_2019,lorenz2021qnlp}. The parameterised quantum circuit corresponding to ``Alice generates language`` is given in figure \ref{fig:qiskit}.

\subsection{Sentence Classification}
\label{sec:sentence_classification}
Before we can present our sentence generation algorithm we must first explain how sentence classification can be performed on near-term quantum devices. What we outline here is a step-by-step overview for solving the following task: Given a dataset $\Gamma$ of sentences, each of which belongs to one of $k$ possible topics, train a classifier that can correctly determine the topic of further unseen sentences (provided the unseen sentences are also about one of the $k$ possible topics). This section mostly follows \citet{lorenz2021qnlp}, although we modify the algorithm to perform multi-class rather than binary sentence classification.

\begin{enumerate}
    \item Each sentence $S \in \Gamma$ is converted to a parameterised quantum circuit $C_S$ using the techniques discussed in the previous section. Note that some parameters may be shared between quantum circuits corresponding to different sentences. This occurs when the same words appear in multiple sentences. We set $q_n = 1$, and $q_s = \lceil \log{k} \rceil$, where $q_n$ and $q_s$ are the number of qubits associated to the noun and sentence wire types respectively. Measuring such a circuit yields one of $k$ possible outcomes, each of which we associated with one of the topics in our corpus.
    \item For each sentence $S \in \Gamma$ and each topic $i \in \{0, 1, ..., k-1\}$ we define a binary predicate $L(i, S) \in \{0, 1\}$ and set $L(i, S) = 1$ if and only if sentence $S$ has topic $i$. Moreover, we write $P(i, C_S)$ for the probability of observing outcome $i$ when measuring the final state of a quantum circuit $C_S$. Finally, let $\Omega$ denote the full set of parameters used in all the quantum circuits combined. Our goal is thus to find the optimal $\Omega$ which maximises $P(i, C_S)$ whenever $L(i, S) = 1$. This problem can be solved using classical machine learning techniques, by minimising the categorical cross-entropy loss function below. This is achieved by using the Simultaneous perturbation stochastic approximation (SPSA) algorithm \cite{Spall1998spsa}.
    
    $$C(\Omega) = \Sigma_{S \in \Gamma} L(i, S) . \log P(i, C_S)$$
    
    \item Given an unseen sentence $S \notin \Gamma$ we can now predict its topic as follows: Use the optimal parameters $\Omega$ to create the quantum circuit $C_S$. Measure the final state of $C_S$ obtaining an outcome $i \in \{0, 1, ..., k-1\}$. Output the topic associated with outcome $i$.
\end{enumerate}

\section{Sentence Generation}
\label{section:algorithm}

In this section, we present our hybrid quantum-classical sentence generation algorithm. 

We first discuss the simulated annealing (SA) algorithm for solving combinatorial optimisation problems \cite{Kirkpatrick1983simannealing}. Then, we rigorously formulate our sentence generation task as an optimisation problem and show in detail how a version of SA can be used to efficiently generate and test many candidate sentences until a satisfactory one is found.

\subsection{Simulated Annealing}

An optimisation problem is a problem where a satisfactory solution must be found from a search space of possible solutions. By a satisfactory solution we mean one that maximises (or comes close to maximising) some objective function over the search space. 

Simulated annealing (SA) is a well-known heuristic method for solving optimisation problems. Let $\mathcal{X}$ be a search space, and $f: \mathcal{X} \rightarrow [0, 1]$ be an objective function over that search space. The goal of SA is to find $x \in \mathcal{X}$ which maximises $f(x)$. SA starts by either randomly or heuristically choosing a starting candidate state $x_0 \in \mathcal{X}$. At each step $t$, the algorithm then considers some neighbouring state $x^*$ of the current candidate $x_t$. If $f(x^*) > f(x_t)$ then the algorithm `accepts' $x^*$ by setting $x_{t+1} = x^*$ and beginning a new iteration.  In the event that $x^*$ is not accepted SA simply sets $x_{t+1} = x_t$ and begins a new iteration. Even if $f(x^*) <= f(x_t)$ SA may still accept $x^*$ with some small probability $e^{\frac{f(x^*)-f(x_t)}{T}}$. This is known as the \emph{metropolis} criterion and depends on an annealing temperature $T$. There are many different options available for calculating $T$ at each timestep. Usually this value is set to be high at the start of SA so that $x^*$ has a high acceptance probability. With each iteration, the value of $T$ decreases, allowing SA to converge towards a solution. In this work, we use the \emph{fast simulated annealing} algorithm which sets $T = \frac{T_{i}}{t + 1}$ at each iteration, where $T_i$ is the initial temperature.

Simulated annealing performs well in practice and is guaranteed to converge towards the optimal solution under reasonable assumptions \cite{Granville1994convergence}. Although in the worst-case this convergence may take a prohibitively long amount of time.

\subsection{The Algorithm}
\label{section:alg}

Let us assume that we have trained a multi-class sentence classifier using the techniques discussed in section \ref{sec:sentence_classification}. The sentence generation task we aim to solve is the following: Given as input one of the topics $i \in \{0, 1, ..., k\}$ which the classifier is trained over, produce a sentence with that topic.

 This task can be seen as an optimisation problem where the search space $\mathcal{X}$ consists of all sentences formed from the vocabulary used to train the classifier\footnote{We could even consider the infinite search space of all possible sentences. However, current limitations in quantum hardware mean that solving this more complicated version of the problem is out of scope for the foreseeable future.} . The objective function $f$ can then simply be defined as $f(S) = P(i, C_S)$. Where $C_S$ is the quantum circuit generated using the optimal parameters $\Omega$. As per the discussion in section \ref{sec:sentence_classification} This function is maximal whenever the sentence $S$ has a high probability of being classified with topic $i$. We now outline the procedure for solving this optimisation problem using SA.
 
 \begin{enumerate}
     \item Start by generating a random candidate sentence $s_0$ from our vocabulary. 
     \item At each step $t$ we generate a neighbouring state $s^*$ of $s_t$. This generation proceeds similarly to the word level editing approach of \citet{miao2019cgmh}.  More specifically, let $s_t = [w_1,w_2,...,w_n]$. $s^*$ is generated by randomly performing one of the following editing operations:
     \begin{itemize}
         \item \emph{Insert}: randomly selects a word $w$ and an index $j$ and sets $s^* = [w_1, ... w_{j-1}, w, w_j, ..., w_n]$. 
          \item \emph{Delete}: randomly selects an index $j$ and sets $s^* = [w_1, ... w_{j-1}, w_{j+1}, ..., w_n]$. 
           \item \emph{Replace}: randomly selects a word $w$ and an index $j$ and sets $s^* = [w_1, ... w_{j-1}, w, w_{j+1}, ..., w_n]$. 
     \end{itemize}
     \item Calculate the values $f(s^*) = P(i, C_{s^*})$ and $f(s_t) = P(i, C_{s_t})$ by running the corresponding quantum circuits many times, and building a probability distribution out of the observed outputs. Decide whether to accept $s^*$ or not according to the SA algorithm.
     \item Continue iterating until you find a sentence $s$ that passes a high threshold $\tau$ along the objective function i.e. $f(s) > \tau$. This indicates that the sentence is with high probability about the topic $i$ as required.
 \end{enumerate}

\subsection{Application to Music Composition}
\label{section:music}

Much like how a sentence is composed of words placed side by side, a musical composition can be seen as a sequence of music snippets placed next to each other. Each snippet itself is in turn composed of musical notes, similarly to how a word is composed of letters belonging to an alphabet. 

This similarity was recently exploited in \cite{miranda2021quantum_music} and used to define a musical version of the DisCoCat framework. The authors then used a CFG to generate a dataset of 100 musical compositions for piano. The generated pieces were annotated manually and placed into one of two classes: rhythmic or melodic. This allowed them to train a quantum classifier that distinguishes rhythmic and melodic musical compositions using the techniques of section \ref{sec:sentence_classification}.

By replacing the sentence classifier mentioned in section \ref{section:alg} with the musical classifier described above, we can adapt our SA-based algorithm for the task of generating musical compositions. In the future we will make musical compositions created using this technique available on our project Github repository\footnote{\url{https://github.com/AminKaramlou/QNLG}}. 

\section{Experiments}
\label{section:experiments}

We now define and attempt to solve two simple sentence generation tasks using the algorithm from the previous section. Our source code is available at \url{https://bit.ly/QuantumNLG}.
To the best of our knowledge, the only other algorithm that can solve these tasks using a quantum computer is what we shall refer to as the Random Generation and Testing (RGT) method of \citet{miranda2021quantum_music}. In fact, this algorithm was initially proposed for music composition rather than sentence generation, but it can straightforwardly be adapted to perform the latter task as well. It works by randomly putting words from a vocabulary next to each other, and evaluating the resulting sentence against the objective function we defined in section \ref{section:algorithm}, until a satisfactory sentence is found. We will implement sentence generation using RGT and compare its performance with our SA-based algorithm.

We do not perform any comparison with state-of-the-art classical methods for solving NLG tasks since it is clear that such methods could easily outperform our proof-of-concept algorithm.

\subsection{Food vs IT}

For our first task, we use the food vs IT data-set created in \citet{lorenz2021qnlp}. This dataset consists of 130 sentences generated using a simple Context-Free Grammar (CFG). Each sentence is manually labelled as being about one of two possible topics, Food or IT. In \citet{lorenz2021qnlp} a quantum classifier is trained using this dataset according to the techniques discussed in section \ref{sec:sentence_classification}. With the help of this classifier, we can implement and analyse the SA and RGT-based sentence generation algorithms on the Food vs IT dataset.

\subsubsection{Simulation results}

Before performing experiments on real quantum hardware we first run our algorithms on a `classical simulator'. As the name suggests, this is a classical device that simulates the behaviour of a real quantum computer. Of course, it is prohibitively expensive to simulate large quantum systems (otherwise there would be no point in building quantum devices). Fortunately, the quantum circuits we are dealing with in this paper are all very small, and can thus be simulated efficiently. All simulations in this section were performed on a 2019 MacBook Air with 16 GB of memory and a 1.6 GHz Dual-Core Intel Core i5 processor.

As is standard within NLG literature \cite{NLGSurvey} we evaluate the quality of free-form generated sentences using the following two criteria: 
\begin{enumerate}
    \item Correctness: Does the generated sentence have the correct topic? 
    \item Fluency: Is the generated sentence grammatically and semantically correct?
\end{enumerate}

Table \ref{tab:food_simulation} shows the result of using a classical simulator to generate 30 sentences about food. The correctness and fluency of each of these sentences have been determined according to the human judgement of the authors. For instance, the sentence ``man debugs software'' was judged as being fluent but incorrect while the sentence ``tasty person prepares dinner'' was judged as being correct but not fluent. 

\begin{table}[h]
\centering
\begin{tabular}{ c|c|c}

  & RGT   & SA \\
 \hline
 Fluent and Correct & 23  & 22 \\
 Fluent and Not Correct & 0  & 0 \\
 Not Fluent and Correct & 4 & 4\\
 Not Fluent and Not Correct & 3 & 4 \\
 \hline
 Avg No. of guesses & 7.56 & 7.46
\end{tabular}
\caption{Results of using a classical simulator to generate 30 sentences about food (Number of guesses refers to the number of candidate sentences evaluated against the objective function by each algorithm).}
\label{tab:food_simulation}
\end{table}

We can see that both the RGT and SA algorithms have performed similarly in terms of the quality of the produced sentences. This is to be expected given that the acceptance condition for a candidate sentence ($f(s) > \tau$) is the same in both cases. We can also see that the average number of sentences guessed before a valid solution is found is almost the same for both algorithms. This is somewhat surprising, given the more rudimentary nature of RGT compared to SA. We believe the reason for this is the small search space associated with this generation task, as well as the fact that many sentences in this space are actually about food. Thus, RGT has a high likelihood of finding a good sentence in only a few guesses. On the other hand, a poor initial guess in the SA algorithm can be very detrimental in this case, since the algorithm might get stuck in a sub-optimal neighbourhood for a few steps. As we shall see in the news headline generation task, this advantage of RGT quickly disappears when dealing with more complicated search spaces. 
\subsubsection{Quantum hardware results}

We now repeat the experiment above on a real quantum computer, namely IBM's 16 qubit \texttt{ibmq\_guadalupe} device. When performing experiments on real quantum hardware, it is important to remember that measuring the final state of a quantum circuit will cause this state to collapse to one of the basis states. This means that the only way we can calculate the probabilities $P(i, C_s)$ needed in step 3 of our generation algorithm is to run and measure the circuit $C_s$repeatedly and create a probability distribution of the observed outcomes. The total number of times a quantum circuit is run in this way is referred to as the number of \emph{shots}. In our case, we ran each circuit for 100000 shots. In the ideal case, results from real quantum hardware will be equivalent to those of simulations. However, imperfections in current prototype devices will lead to sub-optimal performance. The results can therefore be used to benchmark the capacity of current devices for applications of this type.

Table \ref{tab:food_hardware} shows the results of using both the RGT and SA algorithms on real quantum hardware in order to generate 10 sentences about food. Interestingly, these results are very similar to the ones obtained using classical simulators in the previous section. This suggests that our algorithms are potentially robust against the inherent noisiness and imperfections of the current generation of quantum computers. We will aim to test this hypothesis further with more extensive future experimentation. 

\begin{table}[h]
\centering
\begin{tabular}{ c|c|c}

  & RGT   & SA \\
 \hline
 Fluent and Correct & 7  & 7 \\
 Fluent and Not Correct & 0  & 0 \\
 Not Fluent and Correct & 2 & 1\\
 Not Fluent and Not Correct & 1 & 2 \\
 \hline
 Avg No. of guesses & 8.4 & 8.5
\end{tabular}
\caption{Results of using the 16 qubit \texttt{ibmq\_guadalupe} quantum computer to generate 10 sentences about food.}
\label{tab:food_hardware}
\end{table}

\subsection{News Headlines}

As we have seen both the SA and RGT-based sentence generation algorithms performed fairly well on the Food vs IT dataset. In this section, we will test the behaviour of these algorithms on a more challenging dataset consisting of 105 news headlines. Similarly to \cite{lorenz2021qnlp}, we generated this dataset by using a CFG. The sentences were then manually annotated as belonging to one of four possible news headline topics, \emph{entertainment}, \emph{politics}, \emph{sports}, or \emph{technology}. Compared to the Food vs IT dataset this dataset contains more sentence topics, has a larger vocabulary, and has more complicated CFG production rules. When it comes to sentence generation, this means that there is a much larger search space to consider and that there are fewer acceptable sentences in this search space, making the task significantly more challenging.

Table \ref{tab:politics_simulation} shows the results of using SA and RGT to generate 30 sentences about politics. As expected for this more complex dataset, the average number of guesses before finding a viable candidate is much less when using SA rather than RGT\footnote{Note that we treat timeouts as 500 guesses for the purposes of averaging.}.

\begin{table}[h]
\centering
\begin{tabular}{ c|c|c}

  & RGT   & SA \\
 \hline
 Timeouts & 8 & 0 \\
 \hline
 Fluent and Correct & 1  & 11 \\
 Fluent and Not Correct & 4  & 1 \\
 Not Fluent and Correct & 3 & 5\\
 Not Fluent and Not Correct & 14 & 13 \\
 \hline
 Avg No. of guesses & 201.1 & 40.4
\end{tabular}
\caption{Results of using a classical simulator to generate 30 sentences about politics (Timeout refers to runs of the algorithm that failed to find a suitable sentence after 500 guesses)}
\label{tab:politics_simulation}
\end{table}

\section{Related and Future Work}
\label{section:future_work}

We have presented a proof-of-concept algorithm showing how a simple NLG task can be performed on current quantum devices. The algorithm also works for generating musical compositions. Two pieces of related work are worth pointing out:

\begin{itemize}

\item In \citet{abbaszade2021translation} a hybrid quantum-classical algorithm based on DisCoCat is described for sentence translation, a task which has a language generation component to it. Even though the authors do not provide an implementation, this algorithm is well-suited for experimentation on current quantum hardware, as it relies on Quantum Long Short Term Memory (Q-LSTM) \cite{chen2020qlstm}, a quantum machine learning model that is particularly well-suited for near term devices, due to having a modest requirement on qubit counts and circuit depth.

\item \citet{arya2022music} formulates the task of music composition as a Quadratic Unconstrained Binary Optimisation (QUBO) problem. QUBO problems are particularly well-suited for being solved using adiabatic quantum computation (AQC) \cite{farhi2000quantumadiabatic}. This is an alternative to the circuit-based model we learnt about in section \ref{section:prelims}\footnote{Although both models are equivalent in terms of computational power \cite{aharonov2008adiabatic}.}. \cite{arya2022music} then proceeds to solve this QUBO problem using D-Wave quantum computers and generate musical compositions. In future work, it would be interesting to compare this approach to the RGT and SA algorithms we have discussed here.

\end{itemize} 

We conclude with some thoughts on future research directions. 

Clearly, all the works above are limited by the small size of today's quantum computers. However, several companies have announced plans for building significantly more powerful quantum devices in the next few years (see e.g. \citealp{quantum_roadmap}). These devices will undoubtedly be capable of solving more sophisticated NLG tasks than the ones presented here. Whether or not this will eventually lead to quantum algorithms that outperform today's state-of-the-art classical NLG techniques is a fascinating open question that could have dramatic consequences for the field as a whole. We hope that this work serves as sufficient inspiration for the rest of the community to join us in tackling this question.

A further limitation of our techniques is the fact that DisCoCat, while well-suited for modelling the meaning of sentences, is not capable of modelling the meaning of larger pieces of text. This is problematic when it comes to performing more sophisticated NLG tasks e.g. text summarization, given that these tasks often require the production or manipulation of long passages of text. To alleviate this issue, we could use a recently proposed generalisation of DisCoCat, referred to as the Distributional Compositional Circuit-based (DisCoCirc) model \cite{coecke2021discocirc}. Inspired by how DisCoCat uses the grammatical relationship between words to encode the meaning of a sentence, DisCoCirc uses the relationship between sentences to encode the meaning of an entire passage of text. A potential avenue for future work is thus to use DisCoCirc and create a pipeline similar to what we have seen in sections \ref{sec:sentence_classification} and \ref{section:algorithm} for solving document-level rather than sentence-level NLG tasks. 

%\section{Acknowledgements} JRW acknowledges support from the NCCR SPIN, a National Centre of Competence in Research, funded by the Swiss National Science Foundation (grant number 51NF40-180604).


\bibliography{anthology,custom}
\end{document}